\documentclass[twocolumn,preprintnumbers,nofootinbib,prd,superscriptaddress,groupedaddress,aps]{revtex4-1}

\usepackage{graphicx,amssymb,amsmath,amsthm,amsfonts,epsfig,epsf}
\usepackage[linktocpage]{hyperref}
\usepackage[usenames]{color}
\usepackage{epstopdf}
\usepackage{appendix}
\usepackage{aas_macros}
\usepackage{bm}
\usepackage{dcolumn}
\usepackage[latin1]{inputenc}
\usepackage{latexsym}
\usepackage{rotating}
\usepackage{hyperref}
\usepackage{color}
\usepackage{longtable}
\usepackage{enumerate}
\usepackage{tensor}
\usepackage{mathtools}
\usepackage{url}
\usepackage{multirow}
\def\be{\begin{equation}}
\def\ee{\end{equation}}

\newcommand{\bea}{\begin{eqnarray}}
\newcommand{\eea}{\end{eqnarray}}
\newcommand{\ben}{\begin{enumerate}}
\newcommand{\een}{\end{enumerate}}
\newcommand{\bi}{\begin{itemize}}
\newcommand{\ei}{\end{itemize}}

\newcommand{\beq}{\begin{eqnarray}}
\newcommand{\eeq}{\end{eqnarray}} 
\newcommand{\ba}{\begin{align}}
\newcommand{\ea}{\end{align}}
\newcommand{\bsub}{\begin{subequations} \begin{eqnarray}}
\newcommand{\esub}{\end{eqnarray} \end{subequations}}

\begin{document}

\title{Ultralight scalars and resonances in black-hole physics}
\author{Ryuichi Fujita$^{1}$}
\author{Vitor Cardoso$^{1,2,3}$ }
\affiliation{${^1}$ CENTRA, Departamento de F\'{\i}sica, Instituto Superior T\'ecnico -- IST, Universidade de Lisboa -- UL,
Avenida Rovisco Pais 1, 1049 Lisboa, Portugal}
\affiliation{${^2}$ Perimeter Institute for Theoretical Physics Waterloo, Ontario N2J 2W9, Canada}
\affiliation{${^3}$ Theoretical Physics Department, CERN, CH-1211 Gen\`eve 23, Switzerland}


\begin{abstract}
Ultralight degrees of freedom coupled to matter lead to resonances, which can be excited
when the Compton wavelength of the field equals a dynamical scale in the problem.
For binaries composed of a star orbiting a supermassive black hole, these resonances lead to a smoking-gun effect:
a periastron distance which {\it stalls}, even in the presence of gravitational-wave dissipation.
This effect, also called a {\it floating orbit}, occurs for generic equatorial but eccentric orbits and we argue that finite-size effects are not enough to suppress it.
\end{abstract}

\maketitle

\section{Introduction}
Massive scalars or pseudoscalars are a natural extension of General Relativity and have been used
in the Peccei-Quinn theory or improvements thereof to solve the strong {\it CP} problem, in scalar-tensor modifications of gravity, as dark matter models, and in most cosmological models. They arise as moduli and coupling constants in string theory and we now know that at least one scalar exists in nature, namely the Higgs boson. For reviews, see, for example, Refs.~\cite{Berti:2015itd,Magana:2012}.

Consider a massive scalar field $\varphi$ of mass $m_s=\hbar\mu_s$, coupled nonminimally to matter and described at linear order by 
the Klein-Gordon equation sourced by a scalar charge,
\bea
\left[\square-\mu_s^2\right]\varphi=\alpha {\cal T}\label{EQBD2b}\,.
\eea
The source ${\cal T}$ is the trace of the stress-energy tensor and $\alpha$ describes a scalar charge for matter.
This equation describes a wide range of situations~\cite{Cardoso:2011xi,Yunes:2011aa,Berti:2013gfa,Cardoso:2013fwa}.

Take a dynamical situation for which the source ${\cal T}$ is a periodic function of time and describes, for example, 
a two-body problem of characteristic frequency $\Omega_p$. It is then clear that, for scalars with an ``eigenmode'' 
at $\omega=m\Omega_p$, a resonance will be excited by the periodic motion, with $m$ an integer.
In Refs.~\cite{Cardoso:2011xi,Yunes:2011aa} this was shown to be correct for a system composed of a supermassive, spinning black hole (BH) and a pointlike star, driven by gravitational- and scalar-wave emission. The resonance occurs at $\omega=m\Omega_p \sim \mu_s$ and leads to the following:
\begin{enumerate}
\item[(i)] There is a surprising effect for BHs spinning above a critical, $\mu_s$-dependent threshold.
Because of superradiance~\cite{Brito:2015oca}, matter can hover into ``floating orbits'' for which the
net gravitational energy loss at infinity is entirely provided by the BH's rotational energy.
Orbiting bodies remain floating until they extract sufficient angular momentum from the BH,
or until perturbations or other effects disrupt the orbit.
\item[(ii)] There is a speedup in the inspiralling of the star, when the resonance is nonsuperradiant. In other words,
for slowly spinning BHs, the orbiting star ``sinks'' in, once it comes across the resonance.
\end{enumerate}

Floating is, in this context, a nonperturbative effect due to a resonance between the BH and the scalar field (more precisely, induced by the massive term). Any resonant, nonperturbative phenomenon is extremely (``nonperturbatively'') sensitive to perturbations of the conditions that give rise to resonance. Several effects could suppress the resonances, most notably eccentric orbits, finite-size effects and conservative self-force effects.
The purpose of this work is to compute accurately the resonance details for eccentric orbits, and to estimate
whether finite-size effects can destroy the resonances. We show that neither eccentricity nor finite-size effects
seem able to affect the resonance significantly.

\section{Setup}
We consider a stellar-mass compact object inspiralling into a supermassive BH. For such extreme mass ratio inspirals (EMRIs), generic scalar-tensor theories reduce to massive or massless Brans-Dicke theory~\cite{Yunes:2011aa} 
and the field equations for the scalar field at the first-order perturbation are given by Eq.~(\ref{EQBD2b}). 
Our main results are, to a large extent, independent of the source term on the right-hand side, but for concreteness we focus on source terms of the form
\bea
{\cal T}=\int \frac{d\bar\tau}{\sqrt{-\bar g^{(0)}}}\,m_p\delta^{(4)}\left(x-X(\bar\tau)\right)\,,
\eea
corresponding to the trace of the stress-energy tensor of a point particle with mass $m_p$, where $\bar g^{(0)}$ is the background (Kerr) metric
and $X(\bar\tau)$ is the orbit of the particle.
\subsection{Eccentric orbits in the equatorial plane of a black hole}
\label{sec:geodesic}

EMRIs in the eLISA band are expected to have orbital eccentricities, of the order of
$\sim 0.7$~\cite{Hopman:2005vr}. The orbital inclination of the particle 
would increase (decrease) for prograde (retrograde) orbits due to 
the emission of gravitational waves and the rate of change of the orbital 
inclination is very small~\cite{Shibata:1993yf,Shibata:1994jx,Hughes:1999bq,Hughes:2001jr,Sago:2005fn,Ganz:2007rf,Fujita:2009us,Sago:2015rpa}. 
Thus, it is expected that a typical EMRI has both an orbital 
eccentricity and an orbital inclination. 
In this section, however, and for simplicity, we focus on eccentric orbits on 
the equatorial plane of the Kerr BH and investigate 
if the orbital eccentricity affects resonance. 
We discuss the effect of the orbital inclination on the resonance 
in Sec.~\ref{sec:summary}.

When the particle moves on the equatorial plane of the Kerr BH, 
the equations of motion 
are given by 
\bea
r^4 \left(\frac{dr}{d\tau}\right)^2 &=& R(r),
\label{eq:geodesic_r}\\
r^2 \left(\frac{d\phi}{d\tau}\right) &=& \frac{a}{\Delta}\,P_r(r)-(a\,{\cal E}-{\cal L})\equiv \Phi(r),
\label{eq:geodesic_phi}\\
r^2 \left(\frac{dt}{d\tau}\right) &=& \frac{r^2+a^2}{\Delta}\,P_r(r)-a\,(a\,{\cal E}-{\cal L})\equiv T(r),
\label{eq:geodesic_t}
\eea
where ${\cal E}$ and ${\cal L}$ are the energy and the angular momentum of the particle, 
$R(r)=[P_r(r)]^2-\Delta\,[r^2+(a\,{\cal E}-{\cal L})^2]$, $P_r(r)={\cal E}\,(r^2+a^2)-a\,{\cal L}$ and $\Delta=r^2-2Mr+a^2$. 

Geodesics in the equatorial plane of the BH could be specified by 
the constants of motion $({\cal E},{\cal L})$. For the case of a bound orbit in 
the equatorial plane, one can use another set of orbital parameters $(p,e)$, 
where $p$ is a semilatus rectum and $e$ is an eccentricity. 
Using two turning points of the radial motion, 
the apastron $r_a$ and the periastron $r_p$, $(p,e)$ can be defined through
\bea
r_p=\frac{p}{1+e},\,\,\,r_a=\frac{p}{1-e}.
\eea

Solving $R(r_p)=0$ and $R(r_a)=0$ in Eq.~(\ref{eq:geodesic_r}), 
one can compute the set of the constants of motion $({\cal E},{\cal L})$ in terms of 
a given set of orbital parameters $(p,e)$. 
It is worth noting that for a bound orbit in the equatorial plane 
one can compute two fundamental frequencies, $\Omega_r$ and $\Omega_\phi$, 
as~\cite{Glampedakis:2002ya}
\bea
\Omega_r=2\pi \left(2\,\int_{r_p}^{r_a} dr \frac{dt}{dr}\right)^{-1},\,\,\,
\Omega_\phi
=\frac{\Omega_r}{\pi}\int_{r_p}^{r_a} dr \frac{d\phi}{dr}.
\eea
We also note that the fundamental frequencies can be expressed 
in elliptic integrals~\cite{Schmidt:2002qk,Fujita:2009bp}. 

\subsection{Energy fluxes}\label{sec:flux}
Expanding the scalar field in scalar spheroidal harmonics and Fourier transforming,
\bea
\varphi(t,r)=\sum_{l,m}\int d\omega\,e^{im\phi-i\omega\,t}\frac{X_{lm\omega}}{\sqrt{r^2+a^2}}\,S_{lm}(\theta),
\eea
one can rewrite the Klein Gordon equation in the form
\bea
\left[\frac{d^2}{dr_{*}^2}+V_s\right] X_{lm\omega}(r)=\frac{\Delta}{(r^2+a^2)^{3/2}}T_{lm\omega}, \label{eq:Xlmw}
\eea
where~\cite{Ohashi:1996uz}
\bea
V_s
&=&\left(\omega-\frac{m\,a}{r^2+a^2}\right)^2
-\frac{\Delta}{(r^2+a^2)^4}\biggl[\lambda_s (r^2+a^2)^2\cr
&&+2Mr^3+a^2(r^2-4Mr+a^2)\biggr]-\frac{\mu_s^2\Delta}{r^2+a^2},\nonumber
\eea
$\lambda_s$ is the eigenvalue of the scalar spheroidal harmonics $S_{lm}(\theta)$~\cite{Yunes:2011aa} and 
\bea
T_{lm\omega}
=-\alpha\,m_p\,S^{*}_{lm}\int_{-\infty}^{\infty} dt \frac{e^{i\omega t-im\phi(t)}}{(dt/d\tau)}\delta(r-r(t))\,,
\eea
for eccentric and equatorial orbits, where $S^{*}_{lm}=S^{*}_{lm}(\pi/2)$.

To solve the wave equation, we choose two independent solutions
$X_{lm\omega}^{r_+}$ and $X_{lm\omega}^{\infty}$ of the homogeneous version of Eq.~(\ref{eq:Xlmw}), 
satisfying the boundary conditions~\footnote{The asymptotic behaviors of the radial function $X_{lm\omega}$ for $r\rightarrow\infty$ and $r\rightarrow r_+$ can be found by analyzing that of the potential function, $V_s\rightarrow k_{\infty}^2$ ($V_s\rightarrow k_{+}^2$) for $r\rightarrow \infty$ ($r\rightarrow r_+$). See for example Sec.V of Ref.~\cite{Teukolsky:1973ha}.}~\cite{Cardoso:2011xi,Yunes:2011aa}
\bea
X_{lm\omega}^{\infty,r_+}\sim e^{ik_{\infty,r_+}r_*}\,\,\,{\rm as}\,\,\,
r\rightarrow \infty,r_+,
\eeq
where $k_+=\omega-ma/(2Mr_+)$, $k_\infty=\sqrt{\omega^2-\mu_s^2}$
and $dr_*/dr=(r^2+a^2)/\Delta$. Here we take the Kerr geometry written
in Boyer-Lindquist coordinates, and $r_+$ is the location of the event horizon in
those coordinates.

Equation~(\ref{eq:Xlmw}) can be solved by the Green's function method
\bea
X_{lm\omega}&=&
\frac{X_{lm\omega}^{\infty}}{W}\int_{-\infty}^{r_*} dr'_* T_{lm\omega}(r')\frac{\Delta}{(r'^2+a^2)^{3/2}} X_{lm\omega}^{r_+}(r')\cr
&&+\frac{X_{lm\omega}^{r_+}}{W}\int_{r_*}^{\infty} dr'_* T_{lm\omega}(r')\frac{\Delta}{(r'^2+a^2)^{3/2}}X_{lm\omega}^{\infty}(r'),\nonumber
\eea
where $W$ is the Wronskian of the two homogeneous solutions, $W=X_{lm\omega}^{r_+}dX_{lm\omega}^{\infty}/dr_*-X_{lm\omega}^{\infty}dX_{lm\omega}^{r_+}/dr_*$. 

The inhomogeneous solution has the asymptotic form at infinity as 
\bea
X_{lm\omega}(r\rightarrow\infty)
&=& \frac{e^{ik_{\infty}r_*}}{W}\int_{r_+}^{\infty} dr' T_{lm\omega}(r')\frac{X_{lm\omega}^{r_+}(r')}{(r'^2+a^2)^{1/2}},\cr
&\equiv &\tilde{Z}_{lm\omega}^{\infty}\,e^{ik_{\infty}r_*}\,.
\eea

For eccentric and equatorial orbits, 
the source function $T_{lm\omega}$ has a discrete frequency spectrum and 
hence $\tilde{Z}_{lm\omega}^{\infty}$ is given by~\cite{Glampedakis:2002ya} 
\bea
\tilde{Z}_{lm\omega}^{\infty}
= \sum_{k=-\infty}^{\infty} Z_{lmk}^{\infty} \delta(\omega-\omega_{mk}),
\label{eq:Zinf}
\eea
where $\omega_{mk}=m\Omega_\phi+k\Omega_r$ and 
\bea
Z_{lmk}^{\infty}
= -\frac{\alpha\,\Omega_r\,S^{*}_{lm}}{2\pi\,W} \int_{0}^{2\pi/\Omega_r} dt\,e^{i\omega_{mk}t}\,\frac{e^{-im\phi(t)}}{(dt/d\tau)}\frac{X_{lm\omega}^{r_+}(r(t))}{(r(t)^2+a^2)^{1/2}}.\nonumber
\eea

Similarly, at the horizon,
\bea
X_{lm\omega}(r\rightarrow r_+)
&=& \frac{e^{-ik_+ r_*}}{W}\int_{r_+}^{\infty} dr' T_{lm\omega}(r')\frac{X_{lm\omega}^{\infty}(r')}{(r'^2+a^2)^{1/2}},\cr
&\equiv&\tilde{Z}_{lm\omega}^{r_+}\,e^{-ik_+ r_*},
\eea
where
\bea
\tilde{Z}_{lm\omega}^{r_+}
= \sum_{k=-\infty}^{\infty} Z_{lmk}^{r_+} \delta(\omega-\omega_{mk}),
\eea
and 
\bea
Z_{lmk}^{r_+}=-\frac{\alpha\,\Omega_r\,S^{*}_{lm}}{2\pi\,W} \int_{0}^{2\pi/\Omega_r} dt\,e^{i\omega_{mk}t}\,\frac{e^{-im\phi(t)}}{(dt/d\tau)}\frac{X_{lm\omega}^{\infty}(r(t))}{(r(t)^2+a^2)^{1/2}}\,.\nonumber
\eea

The scalar energy fluxes through a sphere at infinity and at the horizon are given by 
\bea
\left\langle{dE \over dt}\right\rangle_t^\infty
&=& \sum_{l,m,k} \omega_{mk}\sqrt{\omega_{mk}^2-\mu_s^2}\, |{Z}_{lmk}^{\infty}|^2,\label{eq:dEdt8}\\
\left\langle{dE \over dt}\right\rangle_t^{r_+}
&=&\sum_{l,m,k} \omega_{mk} \left(\omega_{mk}-\frac{ma}{2Mr_+}\right) |{Z}_{lmk}^{r_+}|^2.\label{eq:dEdtH}
\eea

In scalar-tensor theories, 
the equations for gravitational perturbations about the Kerr background 
are reduced to a differential equation for the Weyl scalar $\Psi_4$~\cite{Yunes:2011aa}, which satisfies the Teukolsky equation~\cite{Teukolsky:1973ha}. 
One can separate the differential equation into radial and angular parts 
using the Fourier-harmonic decomposition, 
\bea
\Psi_4=\frac{1}{(r-ia\cos\theta)^{4}}\sum_{l,m}\int d\omega\,e^{im\phi-i\omega\,t}R_{lm\omega}(r)\,S_{lm\omega}(\theta),\nonumber
\eea
where $S_{lm\omega}(\theta)$ is the spin-2 spheroidal harmonics~\cite{Yunes:2011aa}. The separated radial equation takes the form 
\bea
\left[\Delta^2\frac{d}{dr}\left(\frac{1}{\Delta}\frac{d}{dr}\right)+V_g\right] R_{lm\omega}(r)={\cal T}_{lm\omega}, \label{eq:Rlmw}
\eea
where
\bea
V_g=\frac{K^2+4i(r-M)K}{\Delta}-8i\omega\,r-\lambda_g,\nonumber
\eea
$\lambda_g$ is the eigenvalue of the spin-2 spheroidal harmonics $S_{lm\omega}(\theta)$ and ${\cal T}_{lm\omega}$ is the source term derived from the energy-momentum tensor of the point particle. 

From the asymptotic behaviors of the potential function $V_g$ for $r\rightarrow\infty$ and $r\rightarrow r_+$, the asymptotic forms of the solutions of the radial equation are derived as~\cite{Teukolsky:1973ha}
\bea
R_{lm\omega}(r\rightarrow\infty)&\equiv& {\cal \tilde{Z}}_{lm\omega}^{\infty}\Delta^2\,e^{-ik_+ r_*},\cr
R_{lm\omega}(r\rightarrow r_+)&\equiv& {\cal \tilde{Z}}_{lm\omega}^{r_+}r^3\,e^{i\omega r_*}. 
\eea
Since the source term ${\cal T}_{lm\omega}$ becomes discrete in $\omega$ for eccentric and equatorial orbits, ${\cal \tilde{Z}}_{lm\omega}^{\infty}$ and ${\cal \tilde{Z}}_{lm\omega}^{r_+}$ are given by
\bea
{\cal \tilde{Z}}_{lm\omega}^{\infty}&=& \sum_{k=-\infty}^{\infty} {\cal {Z}}_{lmk}^{\infty} \delta(\omega-\omega_{mk}),\cr
{\cal \tilde{Z}}_{lm\omega}^{r_+}&=& \sum_{k=-\infty}^{\infty} {\cal {Z}}_{lmk}^{r_+} \delta(\omega-\omega_{mk}).
\eea

The gravitational energy fluxes through a sphere at infinity and at the horizon are given by 
\bea
\left\langle{dE \over dt}\right\rangle_t^\infty
&=& \sum_{l,m,k} \frac{1}{4\pi\omega_{mk}^2}\, |{\cal Z}_{lmk}^{\infty}|^2,\\
\left\langle{dE \over dt}\right\rangle_t^{r_+}
&=&\sum_{l,m,k} \frac{\alpha_{lm}(\omega_{mk})}{4\pi\omega_{mk}^2} |{\cal Z}_{lmk}^{r_+}|^2,
\eea
where 
\bea
\alpha_{\ell m}(\omega) =
\frac{256(2Mr_+)^5\,k_+\,(k_+^2+4\tilde\epsilon^2)
(k_+^2+16\tilde\epsilon^2) \omega^3}
{|{\cal C}_{S}|^2}, \nonumber
\eea
with $\tilde\epsilon=\sqrt{M^2-a^2}/(4Mr_{+})$ and 
${\cal C}_{S}$ is the Starobinsky constant given
by~\cite{Teukolsky:1974yv}
\bea
|{\cal C}_{S}|^2 &=&
\left[ (\lambda_g+2)^2 + 4a\omega m - 4a^2\omega^2 \right]
\left[ \lambda_g^2 + 36a\omega m -36 a^2\omega^2 \right]\cr 
&& + (2\lambda_g+3)(96a^2\omega^2 - 48a\omega m)
+ 144\omega^2(M^2-a^2).\nonumber
\eea

\subsection{Orbital evolution}

As explained in Sec.~\ref{sec:geodesic}, for equatorial orbits 
the constants of motion of the particle, $({\cal E},{\cal L})$, depend on a semilatus rectum $p$ and an orbital eccentricity $e$, i.e. ${\cal E}={\cal E}(p,e)$ and ${\cal L}={\cal L}(p,e)$. The orbital evolution of the particle $dp/dt$ and $de/dt$ can be estimated by the following relation:
\bea
\frac{d{\cal E}}{dt}=\frac{\partial {\cal E}}{\partial p}\frac{dp}{dt}+\frac{\partial {\cal E}}{\partial e}\frac{de}{dt},\,\,\,
\frac{d{\cal L}}{dt}=\frac{\partial {\cal L}}{\partial p}\frac{dp}{dt}+\frac{\partial {\cal L}}{\partial e}\frac{de}{dt}. 
\eea

Using balance arguments for the energy and the angular momentum, one can compute the rate of change of the constants of motion $d{\cal E}/dt$ and $d{\cal L}/dt$ due to the gravitational and the scalar fluxes, which are computed in Sec.~\ref{sec:flux}. Inverting the above equations one can find $dp/dt$ and $de/dt$. For circular orbits, $dp/dt=(\partial {\cal E}/\partial p)^{-1}\,d{\cal E}/dt$ since $de/dt=0$. For sufficiently large $p$, the gravitational fluxes reduce the orbital radius and the orbital eccentricity. However, the orbital eccentricity can increase ($de/dt>0$) near the last stable orbit (LSO), which is the boundary between stable and unstable orbits~\cite{Tanaka:1993pu,Cutler:1994pb,Kennefick:1998ab,Glampedakis:2002ya,Fujita:2009us}. Thus, while critical orbits such that $de/dt=0$ near the LSO are possible, the condition $dp/dt=0$ is harder to meet. Floating orbits, with $dp/dt=de/dt=0$, might be possible if $d{\cal E}/dt=0$ and $d{\cal L}/dt=0$ when the gravitational fluxes are entirely compensated by the scalar fluxes due to superradiance for a wide range of scalar-field masses~\cite{Cardoso:2011xi,Yunes:2011aa}. 

\section{Results}
%
\begin{table}[htb]
    \caption{Location of the resonance and corresponding height of the scalar flux $\left\langle{dE/dt}\right\rangle_{lmk}^{r_+}$ normalized by $\alpha^2 M^2\mu^2$ for $q=0.99$, $l=m=1$ and $n=0$. 
    The point particle is orbiting on a circular geodesic. Note that the location (height) of the resonance agrees with that in Table I of Ref.~\cite{Cardoso:2011xi} and Table II of Ref.~\cite{Yunes:2011aa} around seven (four) decimal places.} \label{tab:dEdtH_e0}
    \begin{tabular}{cccc}
      \hline\hline
      $\mu_sM$ & $p/M$ & $\left\langle{dE/dt}\right\rangle_{lmk}^{r_+}$\\
      \hline
      $0.35$& $1.528190558897$ & $-0.05224$ \\
      $0.3$&  $1.785039383463$ & $-0.07522$ \\
      $0.25$& $2.100162438607$ & $-0.09686$ \\
      $0.2$&  $2.535003020406$ & $-0.11998$ \\
      $0.15$& $3.189395564600$ & $-0.14673$ \\
      $0.1$&  $4.334005383932$ & $-0.18276$ \\
      $0.01$& $21.40209871393$ & $-0.48819$ \\
      \hline\hline
    \end{tabular}
\end{table}
We have solved the above equations with an independent code and recovered to within numerical accuracy 
the results reported in Refs.~\cite{Cardoso:2011xi} and \cite{Yunes:2011aa} for the case of circular orbits. 
Our numerical code was developed in Ref.~\cite{Fujita:2009us} to
compute the total gravitational energy flux for eccentric orbits; we truncate the mode sum at $l=7$ to compute the flux. 
When computing the scalar energy flux at the horizon for $l=m=1$, 
we sum over the $k$-modes, in order to achieve convergence at a few decimal places.

In Table~\ref{tab:dEdtH_e0} 
we list orbital radius at resonance in a circular orbit 
and the corresponding height of scalar energy flux at the horizon 
for $l=m=1$ and the $n=0$ mode, and we find results from the independent code 
are consistent with that in Table I of Ref.~\cite{Cardoso:2011xi} and 
Table II of Ref.~\cite{Yunes:2011aa}. 
We note that these resonances occur when frequencies of waves are close to 
the mass $\mu_s$ as~\cite{Detweiler:1980uk}
\beq
\omega_{\rm res}=\mu_s \left[1-\left(\frac{\mu_s M}{l+1+n}\right)^2\right]^{1/2},
n=0,1,\dots.
\eeq
We also note that the height of the scalar flux at resonance is almost same 
at least for the first few overtone modes~\cite{Cardoso:2011xi}. 

\subsection{Eccentric orbits}
%
\begin{figure}[htb]
\begin{center}
\includegraphics[width=0.8\linewidth]{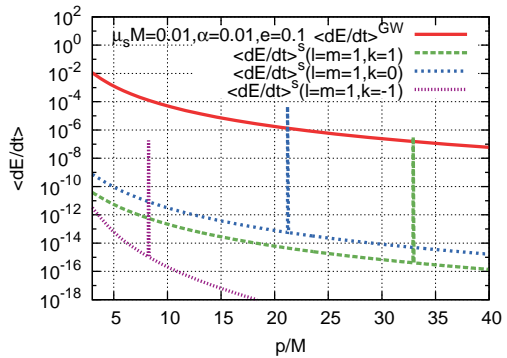}
\caption{Comparison of the total gravitational energy flux with the modal scalar energy flux at the horizon for $l=m=1$, $n=0$, and $k=-1, 0$ and $1$ when $q=0.99$, $\mu_sM=0.01$, $\alpha=0.01$ and $e=0.1$ as a function of $p/M$. For these parameters, floating orbits are possible for $k=0$ and $k=1$ modes. The location and height of the peak and $\alpha_{\rm crit}$ for $l=m=1$ and $n=0$ when $q=0.99$, $\mu_sM=0.01$ and $e=0.1$ are shown in Table \ref{tab:dEdtH}.}\label{fig:dEdtH_l1m1} 
\end{center}
\end{figure}
\begin{table*}[htb]
    \caption{Location of the resonance and corresponding height of the scalar flux $\left\langle{dE/dt}\right\rangle_{lmk}^{r_+}$ normalized by $\alpha^2 M^2\mu^2$ for $q=0.99$, $l=m=1$, $n=0$, and $\mu_s M=0.01$ (left) and $\mu_s M=0.1$ (right), for different eccentricities $e=0.1,0.3,0.5,0.7$. Floating is possible when $\alpha\ge\alpha_{\rm crit}$. Note that, for scalar-tensor theories, 
$\alpha_{\rm crit}$ is well below current observational constraints for $\mu_s$ considered in the table~\cite{Perivolaropoulos:2009ak}, for eccentricities $e=0.1,0.3$ for example.}
\label{tab:dEdtH}
 \begin{center}
   \begin{tabular}{c|ccc|ccc}
      \hline\hline
      \multicolumn{1}{c|}{~} & \multicolumn{3}{c|}{$\mu_sM=0.01$, $e=0.1$} & \multicolumn{3}{c}{$\mu_sM=0.1$, $e=0.1$}\\ 
      $k$ & $p/M$ & $\left\langle{dE/dt}\right\rangle_{lmk}^{r_+}$ & $\alpha_{\rm crit}$ & $p/M$ & $\left\langle{dE/dt}\right\rangle_{lmk}^{r_+}$ & $\alpha_{\rm crit}$\\
      \hline
      $-1$ & $8.242523541258$ & $-1.76\times 10^{-3}$ & $2.7\times 10^{-1}$ & $2.839923283436$ & $-2.88\times 10^{-3}$ & $2.1$\\
      $0$ &  $21.20401320483$ & $-4.83\times 10^{-1}$ & $1.7\times 10^{-3}$ & $4.305965992343$ & $-1.79\times 10^{-1}$ & $1.2\times 10^{-1}$\\
      $1$ &  $32.91726003574$ & $-3.23\times 10^{-3}$ & $6.9\times 10^{-3}$ & $6.276878840413$ & $-1.47\times 10^{-3}$ & $5.5\times 10^{-1}$\\
      $2$ &  $43.13183105281$ & $-2.90\times 10^{-5}$ & $3.7\times 10^{-2}$ & $8.266969425163$ & $-9.35\times 10^{-6}$ & $3.6$\\
      \hline\hline
      \multicolumn{1}{c|}{~} & \multicolumn{3}{c|}{$\mu_sM=0.01$, $e=0.3$} & \multicolumn{3}{c}{$\mu_sM=0.1$, $e=0.3$}\\ 
      $k$ & $p/M$ & $\left\langle{dE/dt}\right\rangle_{lmk}^{r_+}$ & $\alpha_{\rm crit}$ & $p/M$ & $\left\langle{dE/dt}\right\rangle_{lmk}^{r_+}$ & $\alpha_{\rm crit}$\\
      \hline
      $-1$ & $7.905366743932$ & $-1.74\times 10^{-2}$ & $1.0\times 10^{-1}$ & $2.803173882219$ & $-2.73\times 10^{-2}$ & $7.7\times 10^{-1}$\\
      $0$ &  $19.61872491985$ & $-4.37\times 10^{-1}$ & $2.2\times 10^{-3}$ & $4.082557893784$ & $-1.48\times 10^{-1}$ & $1.5\times 10^{-1}$\\
      $1$ &  $30.32848410888$ & $-2.62\times 10^{-2}$ & $3.1\times 10^{-3}$ & $5.840894042062$ & $-1.29\times 10^{-2}$ & $2.3\times 10^{-1}$\\
      $2$ &  $39.69597854286$ & $-2.05\times 10^{-3}$ & $5.7\times 10^{-3}$ & $7.645226999507$ & $-7.62\times 10^{-4}$ & $5.2\times 10^{-1}$\\
      \hline\hline
      \multicolumn{1}{c|}{~} & \multicolumn{3}{c|}{$\mu_sM=0.01$, $e=0.5$} & \multicolumn{3}{c}{$\mu_sM=0.1$, $e=0.5$}\\ 
      $k$ & $p/M$ & $\left\langle{dE/dt}\right\rangle_{lmk}^{r_+}$ & $\alpha_{\rm crit}$ & $p/M$ & $\left\langle{dE/dt}\right\rangle_{lmk}^{r_+}$ & $\alpha_{\rm crit}$\\
      \hline
      $-2$ & $3.275859034959$ & $-2.35\times 10^{-3}$ & $2.1$ & $2.279085908451$ & $-2.91\times 10^{-2}$ & $1.2$\\
      $-1$ & $7.186228834473$ & $-5.92\times 10^{-2}$ & $7.2\times 10^{-2}$& $2.718468024771$ & $-7.20\times 10^{-2}$ & $5.5\times 10^{-1}$\\
      $0$ &  $16.44444756364$ & $-3.43\times 10^{-1}$ & $4.0\times 10^{-3}$ & $3.642199037923$ & $-7.10\times 10^{-2}$ & $3.1\times 10^{-1}$\\
      $1$ &  $25.14958830109$ & $-5.76\times 10^{-2}$ & $3.4\times 10^{-3}$ & $4.976678396309$ & $-3.15\times 10^{-2}$ & $2.3\times 10^{-1}$\\
      $2$ &  $32.82315053092$ & $-1.19\times 10^{-2}$ & $3.9\times 10^{-3}$ & $6.406282378036$ & $-5.84\times 10^{-3}$ & $3.0\times 10^{-1}$\\
      $3$ &  $39.78070525112$ & $-2.87\times 10^{-3}$ & $5.0\times 10^{-3}$ & $7.796703239499$ & $-1.07\times 10^{-3}$ & $4.4\times 10^{-1}$\\
      $4$ &  $46.22083593805$ & $-7.66\times 10^{-4}$ & $6.6\times 10^{-3}$ & $9.122695729626$ & $-2.13\times 10^{-4}$ & $6.8\times 10^{-1}$\\
      \hline\hline
      \multicolumn{1}{c|}{~} & \multicolumn{3}{c|}{$\mu_sM=0.01$, $e=0.7$} & \multicolumn{3}{c}{$\mu_sM=0.1$, $e=0.7$}\\ 
      $k$ & $p/M$ & $\left\langle{dE/dt}\right\rangle_{lmk}^{r_+}$ & $\alpha_{\rm crit}$ & $p/M$ & $\left\langle{dE/dt}\right\rangle_{lmk}^{r_+}$ & $\alpha_{\rm crit}$\\
      \hline
      $-3$ & $2.619283616278$ & $-8.22\times 10^{-3}$ & $1.8$ & $2.222813399026$ & $-3.19\times 10^{-2}$ & $1.3$\\
      $-2$ & $3.304721553556$ & $-2.39\times 10^{-2}$ & $6.6\times 10^{-1}$ & $2.334121278153$ & $-3.19\times 10^{-2}$ & $1.2$\\
      $-1$ & $5.959339601600$ & $-1.65\times 10^{-1}$ & $6.5\times 10^{-2}$ & $2.562147431415$ & $-1.62\times 10^{-2}$ & $1.4$\\
      $0$ &  $11.67081477895$ & $-1.88\times 10^{-1}$ & $1.2\times 10^{-2}$ & $3.011985219663$ & $-3.10\times 10^{-3}$ & $2.2$\\
      $1$ &  $17.37784942139$ & $-7.11\times 10^{-2}$ & $7.1\times 10^{-3}$ & $3.722402383987$ & $-2.04\times 10^{-2}$ & $5.5\times 10^{-1}$\\
      $2$ &  $22.51088131787$ & $-2.74\times 10^{-2}$ & $6.0\times 10^{-3}$ & $4.575213626904$ & $-1.69\times 10^{-2}$ & $3.8\times 10^{-1}$\\
      $3$ &  $27.19777750961$ & $-1.18\times 10^{-2}$ & $5.7\times 10^{-3}$ & $5.456882888116$ & $-7.46\times 10^{-3}$ & $3.7\times 10^{-1}$\\
      $4$ &  $31.55015003696$ & $-5.50\times 10^{-3}$ & $5.8\times 10^{-3}$ & $6.322137823944$ & $-3.02\times 10^{-3}$ & $4.1\times 10^{-1}$\\
      $5$ &  $35.64267125415$ & $-2.74\times 10^{-3}$ & $6.1\times 10^{-3}$ & $7.158618790970$ & $-1.25\times 10^{-3}$ & $4.8\times 10^{-1}$\\
      $6$ &  $39.52621783532$ & $-1.43\times 10^{-3}$ & $6.5\times 10^{-3}$ & $7.964837713480$ & $-5.37\times 10^{-4}$ & $5.6\times 10^{-1}$\\
      $7$ &  $43.23691276357$ & $-7.69\times 10^{-4}$ & $7.1\times 10^{-3}$ & $8.742549001569$ & $-2.41\times 10^{-4}$ & $6.7\times 10^{-1}$\\
      $8$ &  $46.80140470393$ & $-4.26\times 10^{-4}$ & $7.8\times 10^{-3}$ & $9.494295209388$ & $-1.12\times 10^{-4}$ & $8.0\times 10^{-1}$\\
      $9$ &  $50.23999749459$ & $-2.41\times 10^{-4}$ & $8.7\times 10^{-3}$ & $10.22262002496$ & $-5.35\times 10^{-5}$ & $9.6\times 10^{-1}$\\
      \hline\hline
    \end{tabular}
  \end{center}
\end{table*}
The possibility of the existence of floating orbits, i.e., orbits for which the evolution is dominated by superradiance was demonstrated for circular orbits~\cite{Cardoso:2011xi}.
It was also argued, but not calculated, that a small eccentricity would not affect the results.

In this section, we consider the case of eccentric orbits on the equatorial plane of the BH to examine if a floating is possible for massive scalar field. 
As a code check to compute the scalar energy flux for eccentric orbits, 
we have compared the energy flux with the one at the leading post-Newtonian 
order in Ref.~\cite{Will_Zaglauer:1989}, and found our results are consistent
for the case of eccentric orbits and the massless scalar field. 

In Fig.~\ref{fig:dEdtH_l1m1}, we compare the total gravitational energy flux
with the modal scalar energy flux at the horizon for $l=m=1$, $n=0$, and $k=-1, 0$ and $1$ when $q=0.99$, 
$\mu_s M=0.01$, $\alpha=0.01$ and $e=0.1$. 
In the figure, we find three peaks in the scalar energy flux at the horizon 
corresponding to $k=-1$, $0$ and $1$ modes from left to right, and 
find that floating would be possible for $k=0$ and $1$ modes. 
In Table~\ref{tab:dEdtH}, 
we also show the location, the height 
of the peak and the critical value of $\alpha$ for a floating orbit to be possible. We fix $l=m=1$, $n=0$, $q=0.99$ but vary
$\mu_s M$, the $k$-mode and the orbital eccentricity. 
We note that $\alpha_{\rm crit}$ in the table is smaller than current bounds
on $\alpha$ for most of cases, and hence the orbital eccentricity
would not reduce the resonance effect significantly. 

\subsection{Finite-size effects on resonance}

In this section, we would like to estimate finite-size effects on these superradiant resonances.
A proper treatment is too complex and outside the scope of our work.
Instead, we will estimate finite-size effects by modeling an extended body as made up of a collection of pointlike
noninteracting particles. Such simplification is not new~\cite{Haugan:1982fb,Shapiro:1982,Petrich:1985,Nakamura:1987zz}; however,
one needs to exercise extra care in this context. If the swarm of particles is to describe a compact object, then one needs to 
impose that the radial and angular extent is kept finite at all times. There are at least two different ways to realize this idea and restriction.

In Sec.~\ref{sec:circular_arc}, we deal with particles 
on a circular orbit. These particles mimic a body which is extended in the angular direction but pointlike in the radial direction.
In Sec.~\ref{sec:isofrequency}, we drop the assumption that the particles composing the body are on a circular orbit and consider instead particles in eccentric orbits around the equatorial plane of a Kerr BH.
Generically, the proper distance between these two particles varies substantially, and the swarm of particles would not mimic a compact, rigid body. However, for particles with the same fundamental frequencies and (slightly) different 
orbital parameters, i.e. in ``isofrequency pairing''~\cite{WBS2013}, then the proper distance between the particles is finite and varies only slightly and in a controlled way as time goes by. These are therefore a good model also for an extended body.

\subsubsection{Particles in a circular orbit}\label{sec:circular_arc}
Let us first start with a body which is pointlike in the radial direction and composed of $2N+1$ identical pointlike particles of mass $\mu/(2\,N+1)$ in a circular orbit around a Kerr BH of mass $M$. 
We set the orbit of the $j$th particle as 
\bea
\phi_j(t)=\Omega_\phi\,t+\frac{j}{2\,N+1}\delta,\,\,\,j=0,\pm 1,\pm 2,...,\pm N\,,
\eea
where $\Omega_\phi={\sqrt{M}}/({a\sqrt{M}+r_0^{3/2}})$. We choose $\delta=\arcsin(\mu/r_0)$, in order
to model the typical size of a stellar-mass compact star in a circular orbit around a Kerr BH. 
Note in the limit $N\rightarrow\infty$ 
the group of the particles represents a circular arc 
with the angular size $\delta$. 
(For the case $\delta=2\pi$, i.e. a ring, 
see Sec.~6.1 of Part~II in Ref.~\cite{Nakamura:1987zz}.) 

\begin{figure}[ht]
\begin{center}
\includegraphics[width=0.48\linewidth]{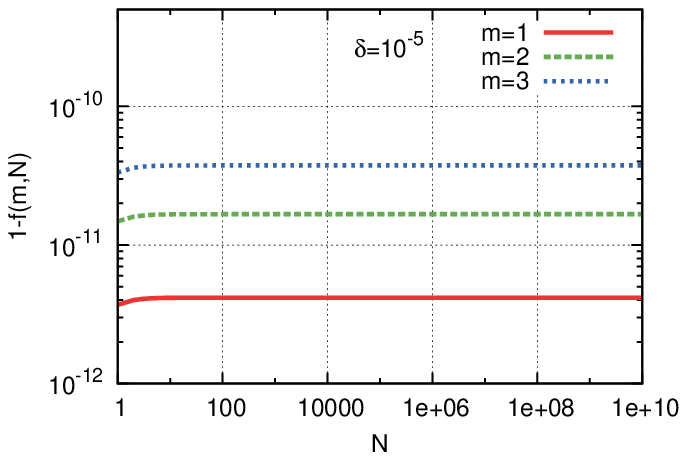}
\includegraphics[width=0.48\linewidth]{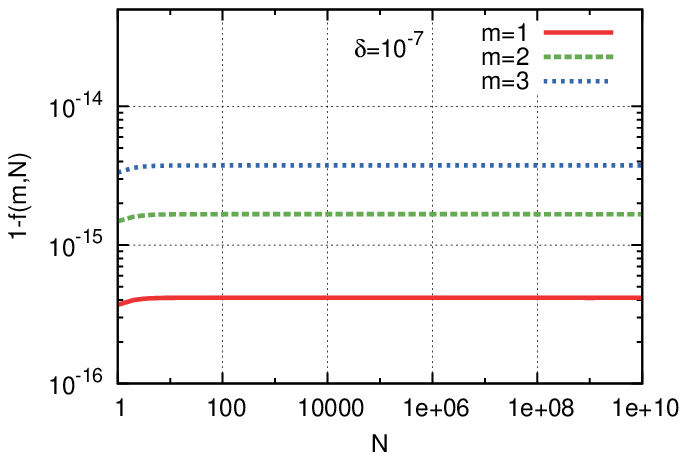}
\caption{Plots for $1-f(m,N)$ when $\delta=10^{-5}$ and $10^{-7}$ from left to right. Finite-size effects along the $\phi$ direction are, for all purposes, negligible since $1-f(m,N)<10^{-10}$ for $m\le 3$, $N\le 10^{10}$ and $\delta\le 10^{-5}$. Note that $f(m,N)$ approaches $1$ if one of $m$, $N$, and $\delta$ approaches zero.}\label{fig:finite_circ}
\end{center}
\end{figure}

The source term of the field equations takes the form
\bea
T_{\ell m\omega}=T_{\ell m\omega}^{(0)}\,f(m,N),
\eea
where 
\bea
f(m,N)
&=&\frac{1}{2\,N+1}\sum_{j=-N}^{j=N}\,e^{i\,m\,\frac{j}{2\,N+1}\,\delta},\cr
&=& \frac{1}{2\,N+1}\left[1+2\,\sum_{j=1}^{j=N}\cos\left(\frac{j}{2\,N+1}\,m\,\delta\right)\right],\cr 
&\le& 1,
\eea
and $T_{\ell m\omega}^{(0)}$ is the source term of a single particle.

The condition $f(m,N)\le 1$ means that in general there are phase cancellations when we take into account the finite-size effects by using a group of particles in a circular arc. In other words,
for a given total mass of the object, point particles radiate the most~\cite{Haugan:1982fb,Shapiro:1982,Petrich:1985,Nakamura:1987zz}.
Note, however, that the cancellation is in general very small 
since $\delta=\arcsin[(\mu/M)(M/r_0)]\ll 1$. Figure~\ref{fig:finite_circ} shows numerical values for $1-f(m,N)$ for $\delta=10^{-5},\,10^{-7}$. We find that $1-f(m,N)<10^{-10}$ for $m\le 3$, $N\le 10^{10}$ and $\delta\le 10^{-5}$.
Moreover, the cancellation affects both the scalar and gravitational channels equally.
Thus, the suppression is under control, and this type of finite-size effects cannot suppress floating
or the resonances.

\subsubsection{Particles in isofrequency orbits}\label{sec:isofrequency}
%
\begin{figure}[ht]
\begin{center}
\includegraphics[width=0.98\linewidth]{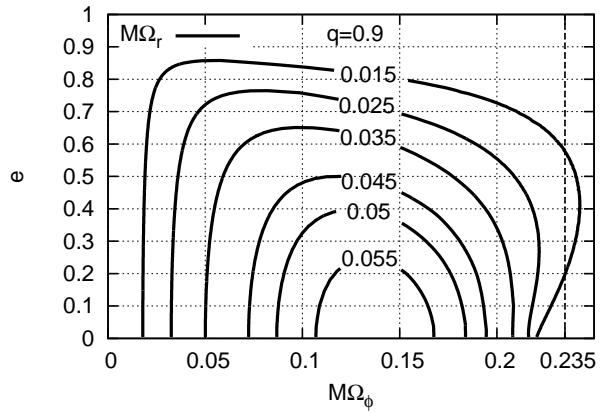}
\caption{Contours for $\Omega_r$ in the $(\Omega_\phi,e)$ parameter space for $q=0.9$. There seems to exist isofrequency orbits for $M\Omega_\phi\gtrsim 0.21$. In the figure, the line $M\Omega_\phi=0.235$ intersects the $M\Omega_r=0.015$ curve at $e\sim 0.2$ and $0.58$; the two intersections are therefore an example of isofrequency orbits. More precise values of orbital parameters for isofrequency orbits are shown in Table \ref{tab:isofrequency}.
}\label{fig:wr_q090}
\end{center}
\end{figure}
Let us now consider an object which is of finite extent in both the radial and azimuthal directions. As a simple toy model, take two particles which are in 
isofrequency pairing~\cite{WBS2013} for eccentric orbits around the equatorial plane of a Kerr BH. 

In an isofrequency pairing situation, two {\it different} orbits can 
have the same fundamental frequencies near the last stable orbit, satisfying the conditions,
\bea
r^{(1)}(t)\neq r^{(2)}(t),&&\,\,\,\phi^{(1)}(t)\neq\phi^{(2)}(t),\cr
\Omega_r^{(1)}=\Omega_r^{(2)},&&\,\,\,\Omega_\phi^{(1)}=\Omega_\phi^{(2)}\,,
\eea
where $\Omega_r^{(j)}$ ($\Omega_\phi^{(j)}$) is the radial (azimuthal) 
frequency of the $j$th particle and $j=1,2$. 

\begin{table*}[htb]
        \caption{Examples of isofrequency orbits for $q=0.7,0.9$ and $0.99$. Note $1-p^{(2)}/p^{(1)}\sim 10^{-7}$, $1-e^{(2)}/e^{(1)}\sim 10^{-6}$, $1-\Omega_\phi^{(2)}/\Omega_\phi^{(1)}\sim 10^{-15}$ and $1-\Omega_r^{(2)}/\Omega_r^{(1)}\sim 10^{-13}$. Note also that the high precision of the parameter values is necessary for the orbital frequencies to agree in the 13 digits. This order of accuracy is necessary because, near the separatrix, small changes in the orbital parameters can result in comparatively large changes in the frequencies. The total energy flux refers to expression, and intends to describe the total flux from a particle with fixed mass (and equal to the masses of the individual particles 1 and 2).}\label{tab:isofrequency}%
    \begin{center}
        \begin{tabular}{c|c|cccc}%
        \hline\hline
	 & Particle & $p$ & $e$ & $\Omega_\phi$ & $\Omega_r$\\
	\hline
	\multirow{2}{*}{$q=0.7$} 
         & $1$ & $3.766722938624189$ & $0.2975030800$ & $0.1434056675641245$ & $0.01911124760358406$ \\
	 & $2$ & $3.766723332045377$ & $0.2975033800$ & $0.1434056675641251$ & $0.01911124760358707$		\\
	\hline
	\multirow{2}{*}{$q=0.9$} 
         & $1$ & $2.611725453072030$ & $0.3042415800$ & $0.2262564780998919$ & $0.02295175692922458$ \\
	 & $2$ & $2.611725688880757$ & $0.3042418000$ & $0.2262564780998936$ & $0.02295175692922107$ \\
	\hline
	\multirow{2}{*}{$q=0.99$} 
         & $1$ & $1.589088238938912$ & $0.1500695000$ & $0.3462538810780511$ & $0.03074999999991420$	\\
	 & $2$ & $1.589088366169729$ & $0.1500697000$ & $0.3462538810780511$ & $0.03074999999994144$	\\
	\hline
	\multirow{2}{*}{$q=0.99$} 
         & $1$ & $1.613113257400407$ & $0.2001356000$ & $0.3513484864560137$ & $0.02759999999997237$	\\
	 & $2$ & $1.613113879695717$ & $0.2001364000$ & $0.3513484864560137$ & $0.02760000000000239$	\\
	\hline
	\multirow{2}{*}{$q=0.99$} 
         & $1$ & $1.645757921740747$ & $0.2499991000$ & $0.3581407568703343$ & $0.02372999999999663$	\\
	 & $2$ & $1.645758096747798$ & $0.2499993000$ & $0.3581407568703343$ & $0.02373000000000163$	\\
	\hline
	\multirow{2}{*}{$q=0.99$} 
         & $1$ & $1.651030264567800$ & $0.2569858000$ & $0.3592171900983110$ & $0.02314659423260921$	\\
	 & $2$ & $1.651030530178192$ & $0.2569861000$ & $0.3592171900983119$ & $0.02314659423261100$	\\
        \hline
	\multirow{2}{*}{$q=0.99$} 
         & $1$ & $1.686790961340073$ & $0.3000463000$ & $0.3663311475282926$ & $0.01947000000000027$	\\
	 & $2$ & $1.686791148229671$ & $0.3000465000$ & $0.3663311475282926$ & $0.01946999999999805$	\\
        \hline\hline
        \end{tabular}
    \end{center}
\end{table*}

Figure~\ref{fig:wr_q090} shows contours for $\Omega_r$ 
in the $(\Omega_\phi,e)$ parameter space when $q=0.9$, 
where $e$ is the eccentricity of 
the orbit defined through $r=p/(1+e\,\cos\chi)$ and $\chi$ is 
the relativistic anomaly parameter. 
(See also Figs.~1 and 3 in Ref.~\cite{WBS2013}.) 
This figure suggests that 
one can find isofrequency orbits 
of which the orbital parameters are very similar, $p^{(1)}\sim p^{(2)}$ and 
$e^{(1)}\sim e^{(2)}$,  
in the region close to the curve along which the Jacobian 
matrix of the transformation $(p,e)\leftrightarrow (\Omega_r,\Omega_\phi)$ 
becomes singular. 

Since
\bea
r^{(j)}(t)&=&r^{(j)}(t+n\,T_r),\,\,\, j=1,2,\cr
\phi^{(j)}(t)&=&\Omega_\phi\,t+\Delta\phi^{(j)}(t)\,,
\eea
where $T_r=2\pi/\Omega_r$ and $\Delta\phi^{(j)}(t)$ is 
the oscillating part of $\phi^{(j)}(t)$ with the period $T_r$, 
$r^{(1)}(t)-r^{(2)}(t)$ and $\phi^{(1)}(t)-\phi^{(2)}(t)$ might be 
constant in the time-averaged sense, 
if isofrequency orbits have very similar orbital parameters. 

\begin{figure*}[ht]
\begin{center}
\includegraphics[width=0.32\linewidth]{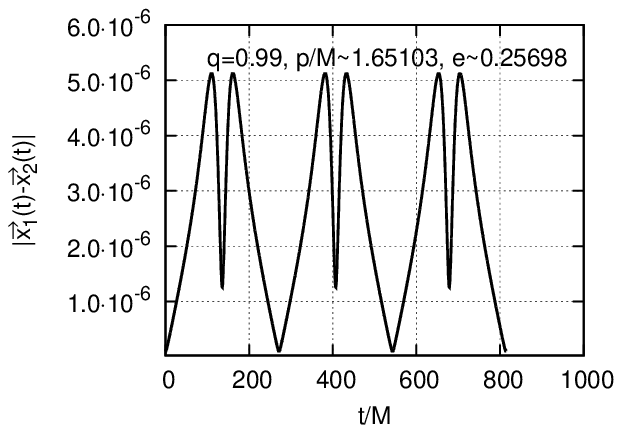}\quad%
\includegraphics[width=0.32\linewidth]{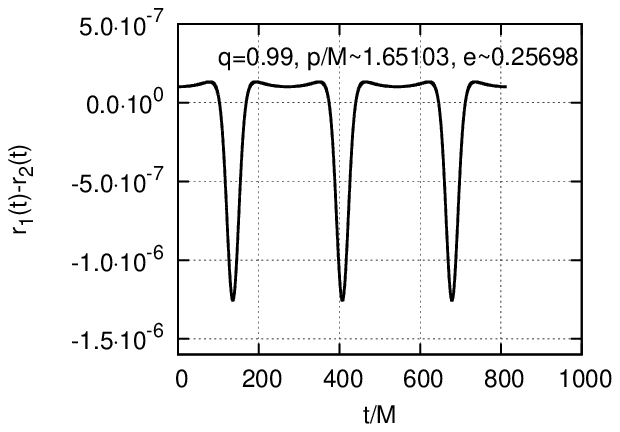}\quad%
\includegraphics[width=0.32\linewidth]{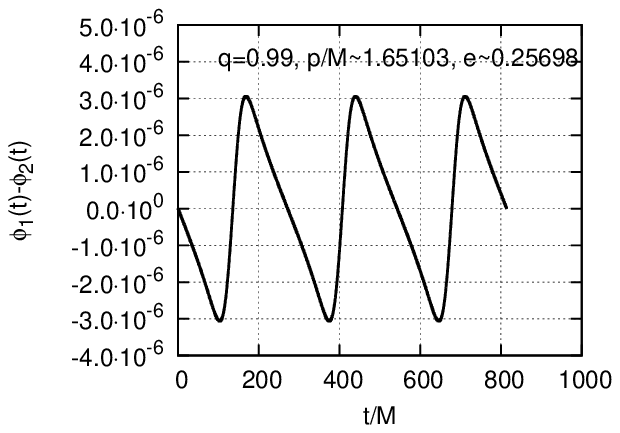}
\caption{Differences in the evolution of $x(t)$ (left), $r(t)$ (middle), and $\phi(t)$ (right), with periods $T_r=2\pi/\Omega_r$, for $q=0.99$ in Table~\ref{tab:isofrequency}. A bounded value of $0<|x^{(1)}(t)-x^{(2)}(t)|\lessapprox 10^{-5}M$ for the difference in position between both particles indicates that the two particles stay in the length scale of mass $\mu\sim 10^{-5}M$.}\label{fig:isofrequency}
\end{center}
\end{figure*}

In Table~\ref{tab:isofrequency}, we show examples of isofrequency
orbits for $q=0.7, 0.9$ and $q=0.99$. 
In the table, we choose the difference 
of orbital parameters for two particles to be as small as 
$1-p^{(2)}/p^{(1)}\sim 10^{-7}$ and $1-e^{(2)}/e^{(1)}\sim 10^{-6}$. 
We still need to show that the distance between the two particles on these orbits is bounded from above by  values that can mimic a compact star.
This study is shown in Fig.~\ref{fig:isofrequency}, where
we show the distance between the two particles as a function of time.
This distance in fact is limited to values bounded from above by the length scale of mass 
$\mu=(\mu/M)M\sim 10^{-5}M$. 
Thus, isofrequency orbits can be used to investigate some of finite-size 
effects from two particles. 

\begin{table*}[htb]
    \caption{
Resonance and corresponding height of the scalar flux $\left\langle{dE/dt}\right\rangle_{lmk}^{r_+}$ normalized by $\alpha^2 M^2\mu^2$ for $q=0.99$ and $l=m=1$, and $k=0$. Orbital parameters $p$ and $e$ are same as those in Table \ref{tab:isofrequency}. The relative difference in the scalar fluxes for two particles in isofrequency orbits, $\left\langle{dE^{(1)}/dt}\right\rangle_{lmk}^{r_+}$ and $\left\langle{dE^{(2)}/dt}\right\rangle_{lmk}^{r_+}$, is comparable to $1-p^{(2)}/p^{(1)}$ and $1-e^{(2)}/e^{(1)}$. 
The averaged scalar flux is the average of the fluxes of the first and the second particles. 
The total scalar flux is computed using the source term 
$\left(T_{\ell m\omega}^{(1)}+T_{\ell m\omega}^{(2)}\right)/2$,
where $T_{\ell m\omega}^{(j)}$ is the source term of the $j$th particle with $j=1,2$. 
The relative difference between the scalar flux for the total and averaged flux is around $10^{-10}$ or better, which is $\sim (1-e^{(2)}/e^{(1)})^2$ and might be understood from Eq.~(\ref{eq:dEdtH}).
\label{tab:dEdtH_isofreq}}
  \begin{center}
    \begin{tabular}{c|c|ccc}
      \hline\hline
      $\mu_sM$ & Particle & $p/M$ & $e$ & $\left\langle{dE/dt}\right\rangle_{lmk}^{r_+}$ \\
      \hline
      \multirow{4}{*}{$0.3524737301$} 
      & $1$ & $1.589088238938912$ & $0.1500695000$ & $-1.495119411\times 10^{-2}$\\ 
      & $2$ & $1.589088366169729$ & $0.1500697000$ & $-1.495114123\times 10^{-2}$\\ 
      & Averaged & $-$ & $-$ &     $-1.495116767\times 10^{-2}$\\
      & Total & $-$ & $-$ & $-1.495116767\times 10^{-2}$\\
      \hline
      \multirow{4}{*}{$0.3578874552$} 
      & $1$ & $1.613113257400407$ & $0.2001356000$ & $-2.048602473\times 10^{-3}$\\ 
      & $2$ & $1.613113879695717$ & $0.2001364000$ & $-2.048527379\times 10^{-3}$\\ 
      & Averaged & $-$ & $-$ &     $-2.048564926\times 10^{-3}$\\
      & Total & $-$ & $-$ & $-2.048564926\times 10^{-3}$\\
      \hline
      \multirow{4}{*}{$0.3651257594$} 
      & $1$ & $1.645757921740747$ & $0.2499991000$ & $-6.931053164\times 10^{-4}$\\ 
      & $2$ & $1.645758096747798$ & $0.2499993000$ & $-6.931127856\times 10^{-4}$\\ 
      & Averaged & $-$ & $-$ &     $-6.931090510\times 10^{-4}$\\
      & Total & $-$ & $-$ & $-6.931090510\times 10^{-4}$\\
      \hline
      \multirow{4}{*}{$0.3662751106$} 
      & $1$ & $1.651030264567800$ & $0.2569858000$ & $-1.134720947\times 10^{-3}$\\ 
      & $2$ & $1.651030530178192$ & $0.2569861000$ & $-1.134733177\times 10^{-3}$\\ 
      & Averaged & $-$ & $-$ &     $-1.134727062\times 10^{-3}$\\
      & Total & $-$ & $-$ & $-1.134727062\times 10^{-3}$\\
      \hline
      \multirow{4}{*}{$0.3738869266$} 
      & $1$ & $1.686790961340073$ & $0.3000463000$ & $-4.014255970\times 10^{-3}$\\ 
      & $2$ & $1.686791148229671$ & $0.3000465000$ & $-4.014260470\times 10^{-3}$\\ 
      & Averaged & $-$ & $-$ &     $-4.014258220\times 10^{-3}$\\
      & Total & $-$ & $-$ & $-4.014258220\times 10^{-3}$\\
      \hline\hline
    \end{tabular}
  \end{center}
\end{table*}
In Table~\ref{tab:dEdtH_isofreq}, we compute the height of 
the resonant scalar flux when $q=0.99$, $l=m=1$, and $k=0$. 
We find that the relative difference in the flux for two particles in 
isofrequency orbits, $\left\langle{dE^{(1)}/dt}\right\rangle_{lmk}^{r_+}$ and $\left\langle{dE^{(2)}/dt}\right\rangle_{lmk}^{r_+}$, 
is comparable to that in the orbital parameters $p$ and $e$, $\sim 10^{-6}$. 
We define two notions of total energy flux, intended to describe an object with fixed total mass.
The ``averaged flux'' is computed as the average of the fluxes of the first and second particles, i.e., $\left(\left\langle{dE^{(1)}/dt}\right\rangle_{lmk}^{r_+}+\left\langle{dE^{(2)}/dt}\right\rangle_{lmk}^{r_+}\right)/2$. 
The ``total'' flux, on the other hand, is computed using the source term
\bea
T_{\ell m\omega}=\frac{T_{\ell m\omega}^{(1)}+T_{\ell m\omega}^{(2)}}{2},
\eea
where $T_{\ell m\omega}^{(j)}$ is the source term of the $j$th particle 
with $j=1,2$. The relative difference in the scalar flux for the total and averaged flux is around $10^{-10}$ or better, which is $\sim (1-e^{(2)}/e^{(1)})^2$ and might be understood from Eq.~(\ref{eq:dEdtH}).

Another issue concerns the extension to $N$ particles in isofrequency 
orbits: there is no guarantee that isofrequency orbits exist for 
$N$-particle systems when $N\ge 3$ (see, for example, Figs.~1 and 3 
in Ref.~\cite{WBS2013}). We may, however, deal with $N$-particle systems 
by considering $\Omega_\phi^{(i)}=\Omega_\phi^{(j)}$, 
$\Omega_r^{(i)}\sim \Omega_r^{(j)}$, $p^{(i)}\sim p^{(j)}$, and $e^{(i)}\sim e^{(j)}$ 
with $0<|x^{(i)}(t)-x^{(j)}(t)|\lessapprox 10^{-5}M$ where $i,j=1,2,...,N$, 
which can be found not only in the region near the LSO but also farther away. 

In fact, one can find such {\it quasi-isofrequency orbits} when $p^{(1)}\le p\le p^{(2)}$. If we define a semilatus rectum $p$ as 
\bea
p^{(i,j)}=p^{(1)}+\frac{2j-1}{2^i}(p^{(2)}-p^{(1)}),
\eea
where $i$ is a positive integer and $1\le j\le 2^{i-1}$, 
we can find corresponding orbital eccentricity $e^{(i,j)}$ which gives 
$e^{(1)}\le e^{(i,j)}\le e^{(2)}$, 
$1-e^{(i,j)}/e^{(i',j')}\lesssim 10^{-6}$, 
$1-\Omega_\phi^{(i,j)}/\Omega_\phi^{(i',j')}\lesssim 10^{-15}$ and 
$1-\Omega_r^{(i,j)}/\Omega_r^{(i',j')}\sim 10^{-13}$. 
If we define semilatus rectum $p$ as $p^{(0,1)}\equiv p^{(1)}$ when $(i,j)=(0,1)$ and $p^{(0,2)}\equiv p^{(2)}$ when $(i,j)=(0,2)$, the total number of particles for $0\le i\le i_{\rm max}$ is given as $N=2^{i_{\rm max}}+1$. 
We note, for the $N$ particle, the difference in semilatus rectums for neighboring particles is given as $(p^{(2)}-p^{(1)})/(N-1)$. 
We compute the scalar flux for the $N$ particle using the source term
\bea
T_{\ell m\omega}=\frac{1}{N}\sum_{i=0}^{i_{\rm max}}\sum_{j} T_{\ell m\omega}^{(i,j)},
\eea
where $T_{\ell m\omega}^{(i,j)}$ is the source term of the particle with the orbital parameters $p^{(i,j)}$ and $e^{(i,j)}$.

\begin{figure}[ht]
\begin{center}
\includegraphics[width=0.98\linewidth]{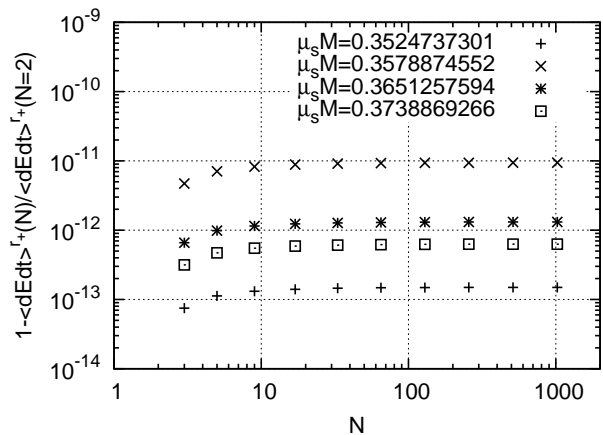}
\caption{The absolute values of the relative difference in the scalar flux for the $N$ particle and that for total are shown as a function of $N$ when $q=0.99$, $l=m=1$, $k=0$ and $0\le i_{\rm max}\le 10$, i.e., $2\le N\le 1025$. The explicit value of the scalar total flux is shown in Table~\ref{tab:dEdtH_isofreq}. 
}\label{fig:diff_dEdtH_N1025}
\end{center}
\end{figure}

In Fig.~\ref{fig:diff_dEdtH_N1025}, the absolute values of the relative difference in the scalar flux for the $N$ particle and that for total are shown as a function of $N$ when $q=0.99$, $l=m=1$, $k=0$ and $0\le i_{\rm max}\le 10$, i.e., $2\le N\le 1025$. The explicit value of the total scalar flux is shown in Table~\ref{tab:dEdtH_isofreq}. 
We choose the total flux as a fiducial value for the comparison since the relative difference in the flux between that and the average estimate with the orbital parameters $p^{(1,1)}$ and $e^{(1,1)}$, which is the center of the $N$ particle along the radial direction, is $\sim 10^{-11}$ or smaller. Since this difference is smaller than the relative difference in the orbital parameters of the $N$ particle $(1-e^{(2)}/e^{(1)})^2$, the finite-size effects along the radial direction are not enough to
suppress floating or the resonance.

%

\section{Conclusion and Discussion}\label{sec:summary}
We have considered a stellar-mass compact star orbiting a supermassive BH in scalar-tensor theories. In Refs.~\cite{Cardoso:2011xi,Yunes:2011aa} 
it was shown that a floating, circular orbit is possible due to 
resonances excited when the orbital period becomes comparable to the mass of the scalar field. In this paper, 
we have extended the analysis in Refs.~\cite{Cardoso:2011xi,Yunes:2011aa} 
to eccentric orbits and to a group of test particles to investigate whether 
resonances due to the coupling of the scalar field to matter in EMRIs 
are affected by the orbital eccentricity 
and finite-size effects of the orbiting star, modeled by the group of particles. 
We have found that these effects do not reduce the resonance significantly. 

From Tables~\ref{tab:dEdtH_e0} and \ref{tab:dEdtH}, 
it is found that, for a given $\mu_s$, scalar fluxes 
for the case of a circular orbit 
are comparable to those for the case of a low eccentric orbit with $k=0$. 
As the orbital eccentricity and $|k|$ increase, however, there appears 
an order of magnitude difference in fluxes for circular and eccentric orbits 
for a given $\mu_s$. This might be understood as follows: 
the scalar energy fluxes are computed through Eqs.~(\ref{eq:dEdt8}) and (\ref{eq:dEdtH}), which require the Wronskian and the integration in time for the asymptotic amplitudes ${Z}_{lmk}^{\infty}$ and ${Z}_{lmk}^{r_+}$. 
Since the Wronskian is computed from the two homogeneous solutions of the Klein-Gordon equation and is independent of the radial coordinate, its value for an eccentric orbit for a given set of $q$, $l$, $m$, $n$, $\mu_s$ and resonant frequency is the same as that of a circular orbit. 
Thus, the main difference in the scalar flux calculation between circular and eccentric orbits 
at resonances for a given $\mu_s$ lies in the integration to find ${Z}_{lmk}^{\infty}$ and ${Z}_{lmk}^{r_+}$. 
Noting the integration becomes zero for circular orbits when $k\neq 0$, we may find that the integration for eccentric orbits when $k\neq 0$ is an order of magnitude smaller than that for circular orbits when $k=0$ for a given $\mu_s$. 
The same argument carries over to inclined orbits, 
and one therefore expects that orbital inclination does not reduce the resonance significantly.

We have also considered a collection of point particles, which were intended to mimic extended bodies and finite-size effects on the resonance. We found that finite-size effects along the azimuthal direction do induce a phase cancellation, but the cancellation is very small and typically unimportant. Since the cancellation affects both the scalar and the gravitational flux equally, we expect that finite-size effects along the azimuthal direction are not enough to suppress floating or resonance. 
We then considered particles in {\it quasi-isofrequency orbits}, that are an extension of isofrequency orbits~\cite{WBS2013}, to take into account finite-size effects along the radial and azimuthal directions. 
Again we found that this type of finite-size effects is very small and is below the relative difference in orbital parameters of particles. Thus we expect that finite-size effects modeled by a collection of particles are not enough to suppress floating or resonance.

The spin of the orbiting particle is another ingredient that should be 
considered to investigate finite-size effects on the resonance. 
The orbits of a spinning particle can be chaotic~\cite{Suzuki:1996gm,Levin:1999zx,Levin:2000md,Hartl:2003da,Cornish:2003uq,Cornish:2003ig,Kiuchi:2004bv,Hartl:2004xr}, and one might not be able to deal with the system in the frequency domain. 
At linear order in the spin of the particle, however, 
it is possible to define frequencies of the orbit and 
spin precession for simple cases~\cite{Mino:1995fm,Tanaka:1996ht}. 
For generic orbits, it is not clear if one can define frequencies of the orbit and 
spin precession, although there still exist three constants of motion~\cite{Gibbons:1993ap}. 
Reference\cite{Ruangsri:2015cvg} suggests solving the evolution of the orbit 
and the spin of the particle using the osculating geodesic method~\cite{Pound:2007th}, making it possible to define the frequencies of the orbit and spin precession at each geodesic~\cite{Ruangsri:2015cvg}. 
In this case, however, the frequencies of the orbit and 
spin precession oscillate in time, and 
the time scale of the oscillation is comparable to that of orbits. 
If the amplitudes of the oscillation in
the frequencies are larger than the width of the resonance in the frequency, 
floating may not be possible 
or may not last long enough to distinguish it from nonfloating orbits.

In summary, we computed the scalar and gravitational energy flux to investigate 
whether floating is possible for eccentric orbits or for extended bodies. 
These fluxes are the time-averaged dissipative part of the first-order 
self-force that would induce deviations
from the geodesic motion at the first order in the mass ratio of the system~\cite{Poisson:2011nh,Barack:2009ux,Pound:2015tma}. 
The conservative self-force is the remaining part of the self-force
which is free from dissipation in the system. 
When one considers the motion taking into account the first-order conservative self-force, 
one may define orbital frequencies in the ``conservative'' effective spacetime~\cite{Detweiler:2002mi}. 
Those frequencies might not coincide with resonant frequencies 
even if orbital frequencies in geodesic motion coincide with 
a resonant frequency. 
It is therefore important to solve the self-force equation in 
scalar-tensor theories~\cite{Zimmerman:2015hua} and to 
compute the self-force to investigate the stability of floating orbits. 
These issues are left for future works. 

\begin{acknowledgments}
We would like to thank Paolo Pani for useful discussions. 
We acknowledge financial support provided under the European Union's H2020 ERC Consolidator Grant ``Matter and strong-field gravity: New frontiers in Einstein's theory'' Grant No. MaGRaTh--646597. 
Research at Perimeter Institute is supported by the Government of Canada through Industry Canada and by the Province of Ontario through the Ministry of Economic Development and 
Innovation.
This project has received funding from the European Union's Horizon 2020 research and innovation program under the Marie Sklodowska-Curie Grant No 690904.
The authors thankfully acknowledge the computer resources, technical expertise and assistance provided by CENTRA/IST and by the Yukawa Institute Computer Facility. Computations were performed at the clusters
``Baltasar-Sete-S\'ois'' and at the Yukawa Institute Computer Facility and were supported by the MaGRaTh--646597 ERC Consolidator Grant.
\end{acknowledgments}


\end{document}